\documentclass[usenatbib]{mnras}
\usepackage[english]{babel}	
\usepackage[utf8]{inputenc}
\usepackage{amsfonts}
\usepackage{amsmath}
\usepackage{siunitx}
\usepackage{blindtext}
\usepackage{supertabular}
\usepackage{enumitem}
\usepackage{mathtools}
\usepackage{soul}
\usepackage{xcolor}					
	\definecolor{nblue}{HTML}{8c8cd9}
	\setulcolor{nblue}
	\setul{}{0.1px}
	
\usepackage[T1]{fontenc}
\usepackage{ae,aecompl}

\usepackage{graphicx}	
\usepackage[config]{caption,subfig}
\usepackage{amsmath}	
\usepackage{amssymb}	
\usepackage{tabu}	

\newcommand{\Msun}{{\rm M_\odot}}	
\newcommand{\defeq}{\vcentcolon=}	
 
\title[Substructure of Abell 2744]{Uncovering substructure with wavelets:\vfill proof of concept using Abell 2744}

\author[J. Schwinn et al.]{%
J. Schwinn$^{1}$\thanks{E-mail: johannes.schwinn@stud.uni-heidelberg.de},
C. M. Baugh$^{2}$,
M. Jauzac$^{3,2,4}$,
M. Bartelmann$^{1}$, 
D. Eckert$^{5}$\\
$^{1}$Universit\"at Heidelberg, Zentrum f\"ur Astronomie, Institut f\"ur Theoretische Astrophysik, Philosophenweg 12, 69120 Heidelberg, Germany\\
$^{2}$Institute for Computational Cosmology, Departement of Physics, University of Durham, South Road, Durham DH1 3LE, U.K.\\
$^{3}$Centre for Extragalactic Astronomy, Department of Physics, Durham University, Durham DH1 3LE, U.K.\\
$^{4}$Astrophysics and Cosmology Research Unit, School of Mathematical Sciences, University of KwaZulu-Natal, Durban 4041, South Africa\\
$^{5}$Max-Planck-Institut f\"ur extraterrestrische Physik, Giessenbachstrasse 1, 85748 Garching, Germany}

\date{Accepted 2018 September 17. Received 2018 September 17; in original form 2018 April 19}

\pubyear{2018}
\begin{document}
\label{firstpage}
\pagerange{\pageref{firstpage}--\pageref{lastpage}}
\maketitle

\begin{abstract}
A recent comparison of the massive galaxy cluster Abell 2744 with the Millennium XXL (MXXL) N-body simulation has hinted at a tension between the observed substructure distribution and the predictions of $\Lambda$CDM. Follow-up investigations indicated that this could be due to the contribution from the host halo and the subhalo finding algorithm used.
To be independent of any subhalo finding algorithm, we therefore investigate the particle data of the MXXL simulation directly. We propose a wavelet based method to detect substructures in 2D mass maps, which treats the simulation and observations equally. Using the same criteria to define a subhalo in observations and simulated data, we find three Abell 2744 analogues in the MXXL simulation. Thus the observations in Abell 2744 are in agreement with the predictions of $\Lambda$CDM. We investigate the reasons for the discrepancy between the results obtained from the {\tt SUBFIND} and full particle data analyses. We find that this is due to incompatible substructure definitions in observations and {\tt SUBFIND}.

\end{abstract}




\section{Introduction}
\begin{figure*}
\centering
\includegraphics[width=1.5\columnwidth]{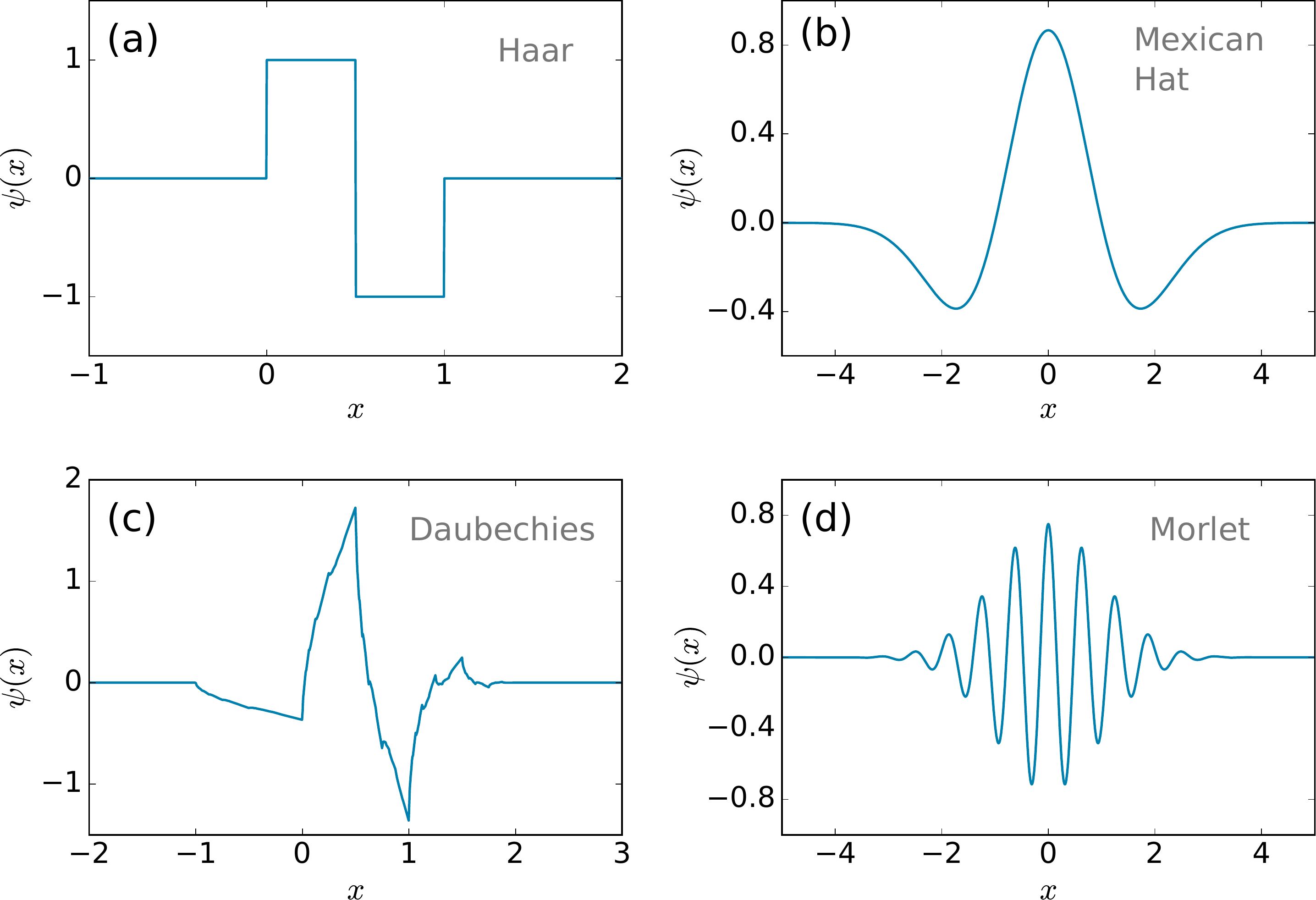}
\caption{\label{fig:wavelets}Examples of four of the most common mother wavelet functions: {\bf (a)} Haar wavelet, {\bf (b)} Mexican Hat wavelet, {\bf (c)}~Daubechies wavelet (with two vanishing moments) and {\bf (d)} real part of the Morlet wavelet. }
\end{figure*}%
Massive galaxy clusters with a high degree of substructure provide an excellent testbed of the current standard model of cosmology \citep{Jauzac2016,Jauzac2017,Natarajan2017,Schwinn2017, Mao2017}. These galaxy clusters comprise more than a thousand galaxies, have masses of more than $10^{15} \Msun$ and can embed several substructures more massive than $10^{14} \Msun$. The standard model of cosmology assumes that the Universe consists of cold dark matter (CDM) in addition to normal baryonic matter and the cosmological constant $\Lambda$ is responsible for the accelerated expansion of the Universe. This model is usually referred to as $\Lambda$CDM. In recent years $\Lambda$CDM has succeeded in describing a variety of observations, such as the fluctuations in the cosmic microwave background (CMB) \citep{Planck2015}, the accelerated expansion of the Universe as measured using type-Ia supernovae \citep{Riess1998,Perlmutter1999}, weak gravitational lensing \citep{Joudaki2017,Koehlinger2017} and the large-scale clustering of galaxies \citep{Cole2005,Alam2016}.

For the purpose of testing $\Lambda$CDM, the galaxy cluster Abell 2744 is ideal. With a mass of $~ 3\times 10^{15}\Msun$ at redshift $z = 0.308$ it is one of the most massive clusters observed in the Universe. Furthermore, it also represents one of the most complex clusters known, with at least seven very massive ($\gtrsim 5 \times 10^{13}\Msun$) merging subhaloes within a distance of $\sim 1$~Mpc from the cluster centre. 

For this reason, we compared in \cite{Jauzac2016} and \cite{Schwinn2017} the distribution of Abell 2744's substructures with the predictions of $\Lambda$CDM using the Millennium XXL (MXXL) N-body simulation \citep{Angulo2012}. Based on the data provided by the structure finding algorithms Friends-of-Friends (FoF) \citep{Davis1985} and {\tt SUBFIND} \citep{Springel2001}, we found a potential tension between the substructure distribution of Abell 2744 and the predictions for a $\Lambda$CDM universe. Following these results, \cite{Mao2017} investigated the high resolution Phoenix cluster simulations \citep{Gao2012} to search for haloes similar to Abell 2744. Using the particle data of the simulation directly, they found one halo with a substructure distribution similar to Abell 2744. They showed further that the host halo contributes a significant amount to the mass measured within 150~kpc apertures around the substructure centres. Furthermore, \cite{Han2017} found that the substructure masses can be significantly underpredicted by {\tt SUBFIND} and thus be in part responsible for the apparent tension with $\Lambda$CDM that is found using subhalo masses alone.

These findings emphasise the need for a consistent identification of substructures in observations and simulations. As \cite{Han2017} show, even among substructure finding algorithms (such as {\tt SUBFIND} or HBT+) different subhalo definitions can lead to disparate results. In addition, a different subhalo definition in observations can lead to further discrepancies. Our aim is therefore to apply the same method to both simulated and observed data to detect substructures and determine their properties. We test the results of \cite{Schwinn2017} without relying on the data provided by FoF \citep[][]{Davis1985} and {\tt SUBFIND} \citep[][]{Springel2001}. Instead we use the particle data of the MXXL simulation directly and investigate if it contains a cluster with properties similar to Abell 2744 (i.e.\ at least seven massive substructures within 1~Mpc). 

 For this task, the wavelet transform is a perfectly suited tool. In the context of detecting substructures within galaxy clusters, it has been introduced by \cite{Escalera1992} and further studies adopt it to analyse a vast number of galaxy clusters \citep[see e.g.][]{Krywult1999, Krywult1999a,Krywult2003,Flin2006}. In a recent study by \cite{Livermore2017}, the wavelet transform has been successfully applied to observations of Abell 2744 in the UV.  Using the wavelet transform, they were able to remove the cluster light in order to detect highly magnified background galaxies. It is our aim to apply a wavelet based method not only to observational data, but to use it for analysing observed and simulated data with one and the same method.

This paper is structured as follows. In Section~\ref{sec:theory}, we introduce the theoretical background of wavelet transforms. In Section~\ref{sec:datasets}, the observational data for Abell 2744 and the simulation data from the MXXL simulation are presented. Section~\ref{sec:massMaps} describes our method, i.e.\ how mass maps are obtained from the simulation and substructures are identified. In Section~\ref{sec:results}, the results of the analysis of observational and simulated data are presented. In Section~\ref{sec:masses}, we discuss the discrepancy between different mass estimates for subhaloes and a possible numerical origin of substructure disruption. We conclude with a summary in Section~\ref{sec:summary}. We would like to stress that throughout the paper we use the terms ``subhalo'' and ``substructure'' interchangeably and abbreviate $M_{200,\mathrm{crit}}$ by $M_{200}$\footnote{The virial mass is usually approximated by $M_{200,\mathrm{crit}}$, which is the mass within a sphere enclosing a mean overdensity of 200 times the critical density of the universe $\rho_\mathrm{crit}$. The radius of this sphere is denoted as $R_{200}$.}. 

\begin{table*}
  \caption{The eight substructures of Abell 2744. Column 1 gives the ID of the substructure, columns 2 and 3 give the position on the sky, column 4 gives the mass within a circular aperture of radius 150~kpc, column 5 gives the significance level of the detection in units of the variance ($\sigma$) in the mass map and column 6 gives the distance of the substructure from the {\it Core}'s brightest cluster galaxy (BCG). The BCG is located at right ascension $\alpha = \ang{3.586259}$ and declination $\delta = \ang{-30.400174}$. This table is based on table 2 from \citet{Jauzac2016}.}
  \centering
    \begin{tabular}{cllccc}
  \hline\hline
  \noalign{\smallskip}
    {\it ID} & R.A. & Dec. & $M(r < 150 \, {\rm kpc})$ & $\sigma$ & $D_{C-S}$ \\
    &(deg)&(deg)&($10^{13} \Msun$)&&(kpc)\\[0.2em] \hline 
    {\it Core} & 3.58626 & -30.40017 & $13.55\pm 0.09$ & 150& - \\
    {\it N} & 3.57666 & -30.35759 & $6.10\pm 0.50$ & 12 & 708.4 \\
    {\it NW} & 3.55310 & -30.37676 & $7.90\pm 0.60$ & 13 & 603.6\\
    ${\it W}_{\it bis}$ & 3.54629 & -30.40332& $5.20\pm 0.60$ & 9 & 565.3\\
    {\it S1} & 3.60412 & -30.37465 & $5.00\pm 0.40$ & 13 & 486.9 \\
    {\it S2} & 3.59895 & -30.35693 & $5.40\pm 0.50$ & 11 & 728.5 \\
    {\it S3} & 3.54151 & -30.37378 & $6.50\pm 0.60$ & 11 & 763.7 \\
    {\it S4} &3.52473 & -30.36958 & $5.50\pm 1.20$ & 5 & 1000.5 \\
    \hline\hline
  \end{tabular}
\label{tab:substructures}
\end{table*}%
\section{Theory - the wavelet transform}
\label{sec:theory}
In signal and data processing, the Fourier transform is a common tool to isolate frequencies of interest, i.e.\ signal contributions of certain length or time scales. This, however, occurs at the price of losing any time or position information about the signal. An analysis in Fourier space is therefore most useful for stationary signals, but has only limited advantage for signals changing with time or position.

An alternative combining the best of both worlds is provided by the \emph{wavelet transform} (WT) \citep{Morlet1982,Daubechies1988, Mallat1989, Meyer1989}. By using a decomposition into wavelets, a signal can be analysed by its wave number without losing positional information. We summarise the main concepts below and refer the reader to the overviews by \cite{Rioul1991} and \cite{Jones2009} or the books of \cite{Daubechies1992} and \cite{Mallat2009} for a more detailed description.

The wavelet decomposition can be used for both discrete signals, using the wavelet series expansion, and continuous signals using the continuous wavelet transform. The continuous wavelet transform (CWT) of a signal $s(x)$ is defined via 
\begin{equation}
\label{eq:cwt}
W_s(a,b) = \frac{1}{\sqrt{a}}\int^{+\infty}_{-\infty} \psi\!\left(\frac{x-b}{a}\right)s(x)\,\mathrm{d}x ,
\end{equation}
where $\psi(x)$ represents the \emph{mother wavelet function}, which is scaled by a parameter $a$ and is shifted by a parameter $b$. In other words, the wavelet transform is given by the convolution of a window function $\psi(x)$ with the signal function $s(x)$. By shifting this window function using the parameter $b$, the positional information of the original signal is preserved. Furthermore, the width of the filter governed by the scaling parameter $a$ introduces a filter scale. The wavelet functions $\psi_{a,b}(x) = \psi\!\left(\frac{x-b}{a}\right)$ are chosen such that they form an orthonormal basis of the $L^2$ (i.e.\ the space of square integrable functions). Furthermore, the mother wavelet function $\psi(x)$ needs to fulfil two conditions: ($i$) it needs to have zero mean
\begin{equation}
\int_{-\infty}^{\infty}\psi(x)\,\mathrm{d} x =0,
\end{equation}
($ii$) it needs to be normalised 
\begin{equation}
||\psi (x)|| = \left[\int_{-\infty}^{\infty}|\psi(x)|^2\, \mathrm{d} x\right]^{1/2} = 1.
\end{equation}
The combination of conditions ($i$) and ($ii$) requires $\psi(x)$ to be a localised, oscillatory function. Furthermore, it can be seen that the prefactor of $1/\sqrt{a}$ in Eq.\,\eqref{eq:cwt} ensures that the scaled wavelet remains normalised according to condition~($ii$). 

There exist many different choices for the mother wavelet in the literature. The most common ones are the Haar wavelet, the Mexican Hat wavelet, the Morlet wavelet, which is a complex valued wavelet, and the family of Daubechies wavelets \citep[see e.g.][]{Daubechies1992,Jones2009,Mallat2009}. We show examples of these four wavelets in Fig.\,\ref{fig:wavelets}. The choice of the mother wavelet depends mainly on the signal analysed. The mother wavelet is chosen such that it best describes the signal that is to be isolated.

\section{Observational and simulated data sets}
\label{sec:datasets}

\subsection{Abell 2744}
\label{subsec:background_A2744}
Abell 2744 is a massive galaxy cluster at redshift $z=0.308$. With a total mass of $\sim 3 \times 10^{15}\,\Msun$ and at least seven substructures with mass $\gtrsim 5 \times 10^{13}\,\Msun$ \citep{Jauzac2016}, Abell 2744 is one of the most massive and most complex galaxy clusters known. For this reason, this cluster has been the subject of a large number of investigations in many wavebands \citep{Merten2011,Owers2011,Eckert2015,Jauzac2015,Medezinski2016,Jauzac2016,Jauzac2017}. 

For our analysis, we use the combined strong and weak lensing mass reconstruction of Abell 2744 from \cite{Jauzac2016}. This reconstruction is based on observations with the \emph{Hubble Space Telescope} (HST) and the \emph{Canada-France-Hawaii Telescope} (CFHT). Full details of the cluster mass reconstruction, including the selection of background galaxies, shape measurements and noise estimation, can be found in \cite{Jauzac2016}.

Using these data, the mass of the cluster within a circular aperture of $R = 1.3$~Mpc was determined as $M(R<1.3\,{\rm Mpc}) = \left( 2.3 \pm 0.1 \right) \times 10^{15}\,{\rm M}_\odot$. Furthermore, the reconstructed mass distribution revealed eight substructures within a distance of 1~Mpc from the cluster centre, all with masses $\gtrsim 5 \times 10^{13}\,\Msun$. For completeness the ID, position on the sky, mass, significance and distance of all eight substructures are listed in Table \ref{tab:substructures}. \cite{Jauzac2016} note that the substructure ${\it W}_{\it bis}$ is probably a background structure projected onto the cluster, since it has a relatively high mass-to-light ratio and the spectroscopic redshifts of galaxies in its vicinity place it behind the cluster. For this reason, we search for a halo within a $\Lambda$CDM universe with at least seven substructures similar to those of Abell 2744.

\subsection{The Millennium XXL Simulation}
\label{subsec:background_MXXL}
As in \cite{Schwinn2017}, we use the Millennium XXL (MXXL) simulation \citep[][]{Angulo2012} to compare our observational results to the $\Lambda$CDM predictions.
The MXXL simulation is the third in the family of Millennium simulations. Using a box size of $3\,h^{-1}{\rm Gpc}$, it was run to investigate structure formation especially on cosmological scales. 
It models dark matter in a $\Lambda$CDM universe with the cosmological parameters set to: $H_0~=~73\,{\rm km\,s^{-1}Mpc^{-1}}$, $\Omega_\Lambda = 0.75$, $\Omega_{\rm m} = \Omega_{\rm dm} + \Omega_{\rm b} = 0.25$, $\Omega_{\rm b} = 0.045$ and  $\sigma_8 = 0.9$. These parameters were chosen such that they are consistent the previous Millennium runs \citep[][]{Springel2005,Boylan-Kolchin2009}. The dark matter fluid is represented by 303 billion particles each having a mass of $m_\mathrm{p} = 8.80 \times 10^{9}\, {\rm M}_{\odot}$.

Gravitationally bound structures are identified at the \emph{halo} level using the Friends-of-Friends (FoF) algorithm \citep[][]{Davis1985}. Within these haloes \emph{subhaloes} are found using
{\tt SUBFIND} \citep[][]{Springel2001}. The FoF algorithm finds haloes by identifying objects built up by connecting particles that are separated less than a given linking length, $b$, which is specified in units of the mean interparticle separation. In the MXXL simulation the linking length was set to $b=0.2$. This ensures that the mean density of FoF haloes corresponds to $\sim 180$ times the critical density (\mbox{$\rho_\mathrm{crit}\defeq 3H^2/(8\pi G)$}) as shown by \citet{More2011}. The {\tt SUBFIND} algorithm identifies substructures by detecting a saddle point in the density profile and checks that the subhalo is gravitationally self-bound.

The properties of all FoF and {\tt SUBFIND} haloes were stored for 64 snapshots ranging from $z = 63$ to 0. However, the position and velocities of the dark matter particles were only stored for four snapshots at redshifts \mbox{$z = 0, 0.24, 1$ and 3}. This data reduction was necessary due to the large amount of storage space needed for one snapshot of full particle data. Our analysis is based on the FoF data sets as well as the full particle data at snapshot 54 ($z = 0.24$), which is the snapshot closest to the redshift of Abell 2744 for which the full particle data available. In a second step, we then compare our findings with the {\tt SUBFIND} data sets. 

\section{Comparing observational and simulated mass maps}
\label{sec:massMaps}
\subsection{Mass maps from the MXXL simulation}
\label{subsec:massMapsMXXL}
Since we aim to perform a comparison of observations and simulations in an as unbiased way as possible, we do not rely on any substructure finding algorithm, but instead we use the MXXL particle data directly to obtain mass maps equivalent to that of Abell 2744. We then compare both the observed and simulated mass maps using the same set of criteria.

In order to obtain projected mass maps for all MXXL haloes with a mass similar to Abell 2744, we select all FoF-haloes at redshift $z = 0.28$ with a mass of $M_{200} \geq 2.0 \times 10^{15} \Msun$. This represents a rather conservative choice, since it corresponds to the lower 3$\sigma$-bound of Abell 2744's mass within an aperture of 1.3~Mpc obtained in \cite{Jauzac2016}. Since this radius is smaller than $R_{200}$, the virial mass of Abell 2744 was estimated in \cite{Schwinn2017} as $M_{200} = 3.3 \pm 0.2 \times 10^{15} \Msun$, which lies well above our lower threshold. We find 209 haloes in the MXXL simulation fulfilling this mass criterion. We then use the particle data of the MXXL simulation at redshift $z=0.24$ (corresponding to the snapshot closest to the redshift of Abell 2744 for which full particle data is available) to create projected mass maps of each halo. We project all particles over a length of 30~Mpc on a $3 \times 3$~Mpc map and bin into pixels of side length $4.55$~kpc. This choice corresponds to the resolution of the mass map obtained for Abell 2744. We have checked that our results are insensitive to the exact choice of the projection length. We obtain for each halo three mass maps using either the $x$-, $y$- or $z$-axis as the line-of-sight. Due to the limited mass resolution of the MXXL simulation with a particle mass of  $m_{p} = 8.80 \times 10^{9}\, {\rm M}_{\odot}$, the mass maps from the simulation are much more coarse-grained than those obtained from observations. In order to correct for this effect, we smooth all mass maps (that of Abell 2744 as well) with a Gaussian filter with a standard deviation of 1.5 pixels ($\sim 6.8$~kpc).

\subsection{Finding substructures}
Our next step is to identify substructures in the mass maps. In order to treat simulated and observed data equally, we do this without using the {\tt SUBFIND} data from the MXXL simulation. The aim is to identify the positions of all peaks on sub-cluster scales in our mass map. The wavelet transform introduced in Section~\ref{sec:theory} provides an excellent tool to extract the mass signal at the scale of interest, without losing the positional information as would be the case with a Fourier transform. We thus use the coefficients of the wavelet transform (WT) to identify significant mass peaks in each mass map. We then define a threshold in the WT coefficients to select only substructures that are as significant as those of Abell 2744. 

While doing so, one needs to keep in mind that the apparent substructure mass consists of two components: ($i$) the background mass distribution of the host halo and ($ii$) on top of that the matter gravitationally bound to the substructure. The WT coefficients depend on both of these components and the host halo boosts the coefficients of the substructures. Therefore a small fluctuation close to the centre can have a higher coefficient than a substructure further away with a higher density peak in comparison to the local background. To avoid this, we fit the mass distribution of the main halo with an NFW density profile \cite{Navarro1996} and subtract its contribution from the mass map before performing the wavelet transform. Since this step removes the central substructure as well, we will add it to the list of detected substructures in the aftermath. Due to the unrelaxed state of Abell 2744 and similar clusters, however, the NFW-profile could potentially provide an insufficient approximation of the main halo's density profile. We therefore verified that this choice only leads to minor changes in the results by applying our method with and without the main halo subtraction (see also Section \ref{subsec:ResultsMXXL}). However, depending on the object being analysed, it might be helpful to run our method in both modes.

\begin{figure*}
\begin{tabular}{cc}
\subfloat[]{\hspace{-0.9em}\includegraphics[width=0.5\linewidth]{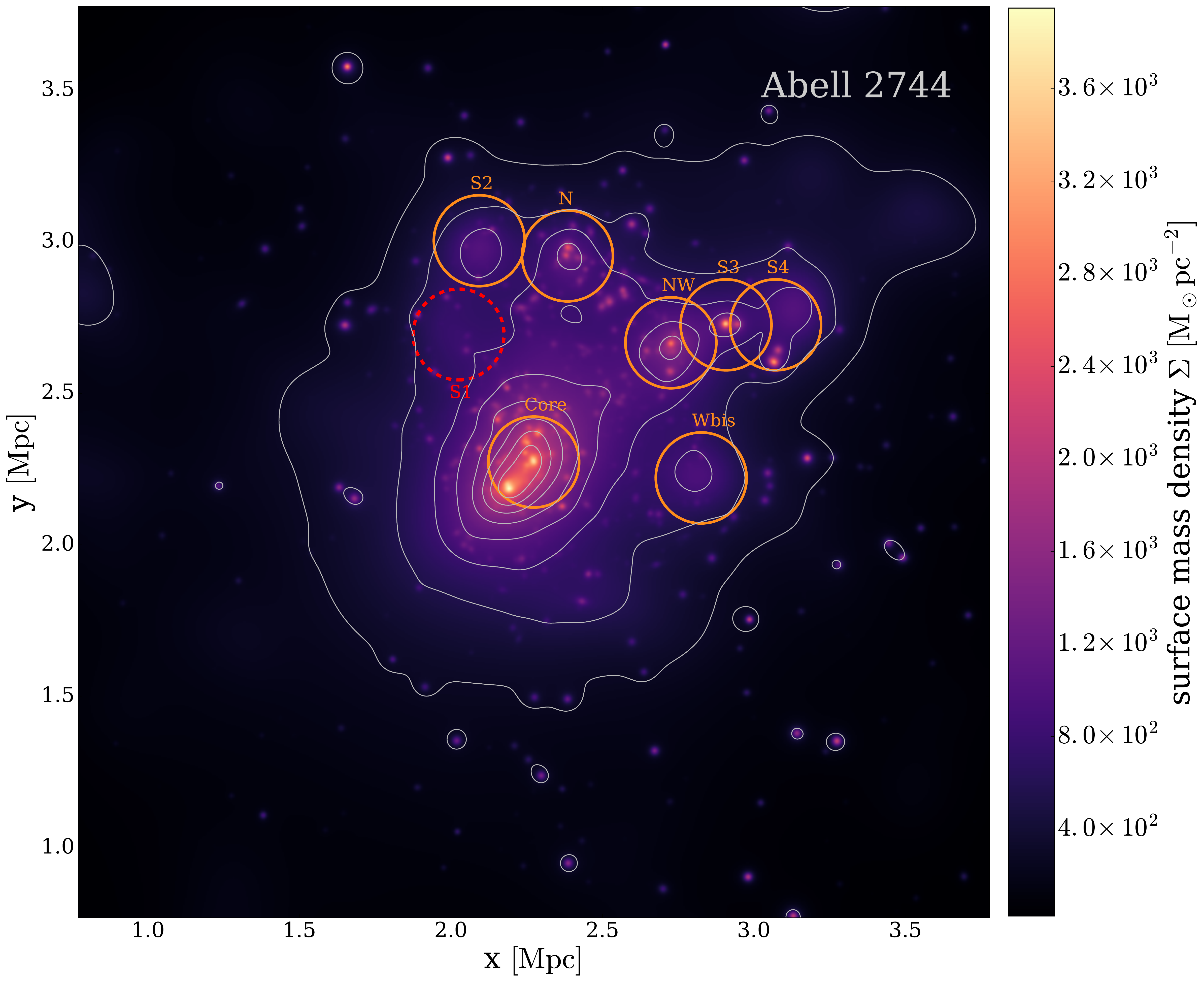}} &
\subfloat[]{\hspace{-0.9em}\includegraphics[width=0.5\linewidth]{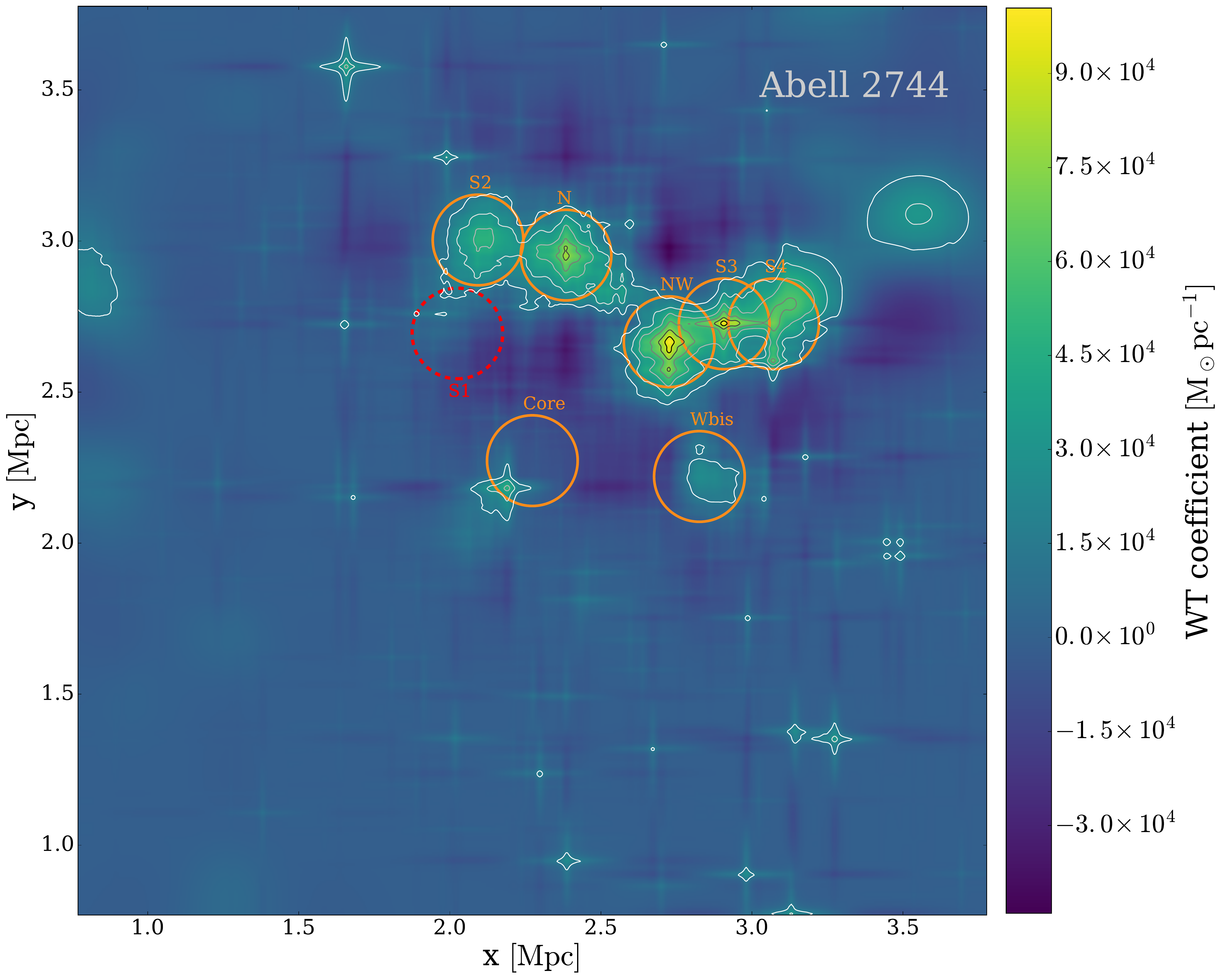}}
\end{tabular}
    \caption{Substructures of Abell 2744 identified automatically by using a wavelet transform. The \emph{left} panel shows the mass map of Abell 2744 where the colour map and contours show the surface density. The \emph{right} panel shows the wavelet transform coefficients computed as described in the text. Substructures fulfilling the defined threshold criteria are marked as orange circles with radius corresponding to $R = 150$~kpc. The \emph{S1} substructure found in \citet{Jauzac2016} is highlighted as a red dashed circle.}
    \label{fig:mapsAbell2744}
\end{figure*}%
The wavelet transform is performed using the 1D continuous wavelet transform module of the {\tt PyWavelet} package\footnote{http://pywavelets.readthedocs.io}. We adopt the Mexican Hat wavelet as the mother wavelet function, since in comparison to all other wavelets it best resembles the shape of the subhaloes we are aiming to extract (see Fig.\,\ref{fig:wavelets}). The 1D wavelet transform is applied row-wise and column-wise to the 2D mass maps. The final WT coefficients are then obtained by taking the arithmetic mean of the row-wise and column-wise coefficients. 

Using this method, a map of WT coefficients is obtained, in which peaks can be detected automatically. We define a quantitative criterion based on the WT coefficient that determines which of the identified substructures we consider to be equally significant as those of Abell 2744. This criterion consists of two parameters: the scale at which the wavelet transform is performed and the threshold for its coefficients, marking the threshold for the significance of the substructure. Our criterion is tailored such that it restores as many of the eight substructures of Abell 2744 found in \cite{Jauzac2016} as significantly as possible. We therefore choose:
\begin{itemize}[leftmargin=1.5em]
\setlength{\itemindent}{-0.5em}
\item[-] a scale of 40~pixels, corresponding to 182~kpc and\vspace{0.5em}
\item[-] a threshold for WT coefficients of $W \geq 2.6 \times 10^{10} \Msun \mathrm{pc^{-1}}$. 
\end{itemize}
These choices maximise the signal-to-noise ratio of Abell 2744's substructures.
This procedure ensures that the substructures found in the simulated mass maps are at least as significant as the substructures found in Abell 2744 and we do not include random fluctuations in our analysis. 

We detect the substructures by selecting all pixels that are at least 5 times above the average WT coefficient of the map. We then select 20 per cent\footnote{This helps to save computational time and prioritises substructures with several pixels above the threshold.} of these pixels randomly and draw a circular aperture around them with a radius of 100~kpc. Since this radius is smaller than the aperture used to determine the mass, it allows us to have slightly overlapping subhalo apertures. In each aperture around the randomly selected pixels, we select the pixel with the largest WT coefficient and centre the aperture on this pixel. We perform this procedure iteratively ten times. This method ensures that all peaks are found and that peaks have a minimal distance of 100~kpc to each other. Peaks closer than that would have significantly overlapping apertures, preventing their masses from being determined independently. We then add the central substructure to the list and remove any other detected substructure within a radius of 150~kpc from the central substructure. As a last step we discard all substructures with an aperture mass $M(R<150~\mathrm{kpc}) < 3 \times 10^{13} \Msun$ or which are at a distance $R > 1.25$~Mpc from the halo centre.

\begin{table*}
  \caption{Properties of substructures found in Abell 2744 and in three MXXL clusters. For all clusters the mass within an aperture of 150~kpc, the distance from the cluster centre and the WT coefficient (wtc) of the centre pixel is given. The mass and distance from the centre for Abell~2744's substructures represent the values found by \citet{Jauzac2016}.}
  \vspace{0.5em}
  \centering
  \resizebox{\textwidth}{!}{%
  \begin{tabular}{c|ccc|c|ccc|ccc|ccc}
    \hline\hline
    \noalign{\smallskip}

    \multicolumn{4}{c}{\emph{Abell 2744}} & \multicolumn{1}{c}{ }&\multicolumn{3}{c}{\emph{halo 37}} &\multicolumn{3}{c}{\emph{halo 95}}&\multicolumn{3}{c}{\emph{halo 114}} \\
    \emph{ID} & $M_{150}$ & $D_{C-S}$ & wtc & \emph{ID} & $M_{150}$ & $D_{C-S}$ & wtc & $M_{150}$ & $D_{C-S}$ & wtc & $M_{150}$ & $D_{C-S}$ & wtc \\    & [$10^{13}\, {\rm M}_\odot$] & [kpc] & [$10^{10}\, {\rm M}_\odot \mathrm{pc}^{-1}$] & & [$10^{13}\, {\rm M}_\odot$] & [kpc] & [$10^{10}\, {\rm M}_\odot \mathrm{pc}^{-1}$] & [$10^{13}\, {\rm M}_\odot$] & [kpc] & [$10^{10}\, {\rm M}_\odot \mathrm{pc}^{-1}$] & [$10^{13}\, {\rm M}_\odot$] & [kpc] & [$10^{10}\, {\rm M}_\odot \mathrm{pc}^{-1}$] \\[0.2em]
\hline
    {\it Core} & $13.55 \pm 0.09$ & - & - & 1 & 12.80 & - & - & 11.40 & - & - & 10.60 & - & - \\
    {\it NW} & $7.90\pm 0.60$ & 604 & 10.00 & 2 & 8.85 & 576 & 17.00 & 8.48 & 524 & 20.30 & 9.50 & 787 & 28.10 \\
    {\it S3} & $6.50\pm 0.60$ & 764 & 9.75 & 3 & 7.35 & 611 & 11.90 & 7.22 & 466 & 11.50 & 9.45 & 171 & 8.91 \\
    {\it N} & $6.10\pm 0.50$ & 708 & 8.04 & 4 & 6.97 & 1182 & 19.40 & 5.48 & 492 & 8.57 & 6.05 & 1182 & 20.10 \\
    {\it S4} & $5.50\pm 1.20$ & 1001 & 6.99 & 5 & 6.73 & 974 & 13.90 & 5.06 & 532 & 5.07 & 4.24 & 842 & 8.47 \\
    {\it S2} & $5.40\pm 0.50$ & 729 & 2.61 & 6 & 6.37 & 902 & 11.00 & 4.11 & 841 & 8.61 & 4.09 & 934 & 7.03 \\
    {\it $W_{bis}$} & $5.20\pm 0.60$ & 565 & 4.91 & 7 & 5.00 & 967 & 6.03 & 3.25 & 793 & 4.46 & 4.00 & 1229 & 5.78 \\
    {\it S1*} & $5.00\pm 0.40$ & 487 & - & 8 & 4.03 & 691 & 5.64 & 3.03 & 997 & 3.48 & 3.54 & 851 & 5.14 \\
    {\it -} & - & - & - & 9 & 3.08 & 1008 & 6.47 & - & - & - & - & - & -\\
    \hline\hline
  \end{tabular}
  }
  \label{tab:substrComparison}
\end{table*}%
\section{Results}
\label{sec:results}
\subsection{Abell 2744} 
Applying our method to the mass map obtained from the combined weak and strong lensing mass reconstruction of \cite{Jauzac2016}, we are able to recover seven of the eight substructures they report. In Fig.\,\ref{fig:mapsAbell2744}, the identified subhaloes are shown in the projected mass map together with a map of the corresponding WT coefficients. However, our algorithm does not identify the \emph{S1} substructure, which was found in both \cite{Medezinski2016} (where it was named \emph{NE}) and \cite{Jauzac2016}. Although this substructure produces a lensing signal with a high significance \citep[13$\sigma$ in][]{Jauzac2016}, its contribution to the mass map is rather modest. Since our detection algorithm is based on the mass of substructures and their scale, it fails to identify this substructure. We test by how far the threshold has to be lowered until \emph{S1} can be recovered. However, we find that once \emph{S1} can be detected, we also pick up several spurious substructures. This does not change either when the wavelet scale is varied in a range between 50 and 200~kpc. For this reason, we do not take $S1$ into account for our comparison with MXXL haloes. We therefore consider a cluster to be like Abell 2744 in terms of its substructure distribution, if at least seven substructures are found that:
\begin{itemize}[leftmargin=1.5em]
\setlength{\itemindent}{-0.5em}
\item[-] are identified by our wavelet transform algorithm with the above defined thresholds,
\item[-] have an aperture mass of at least $M(R<150~\mathrm{kpc}) \geq 3 \times 10^{13} \Msun$, 
\item[-] have a projected distance not greater than 1.25~Mpc from the cluster centre.
\end{itemize}

\begin{figure*}
\begin{tabular}{cc}
\subfloat[]{\hspace{-0.9em}\includegraphics[width=0.45\linewidth]{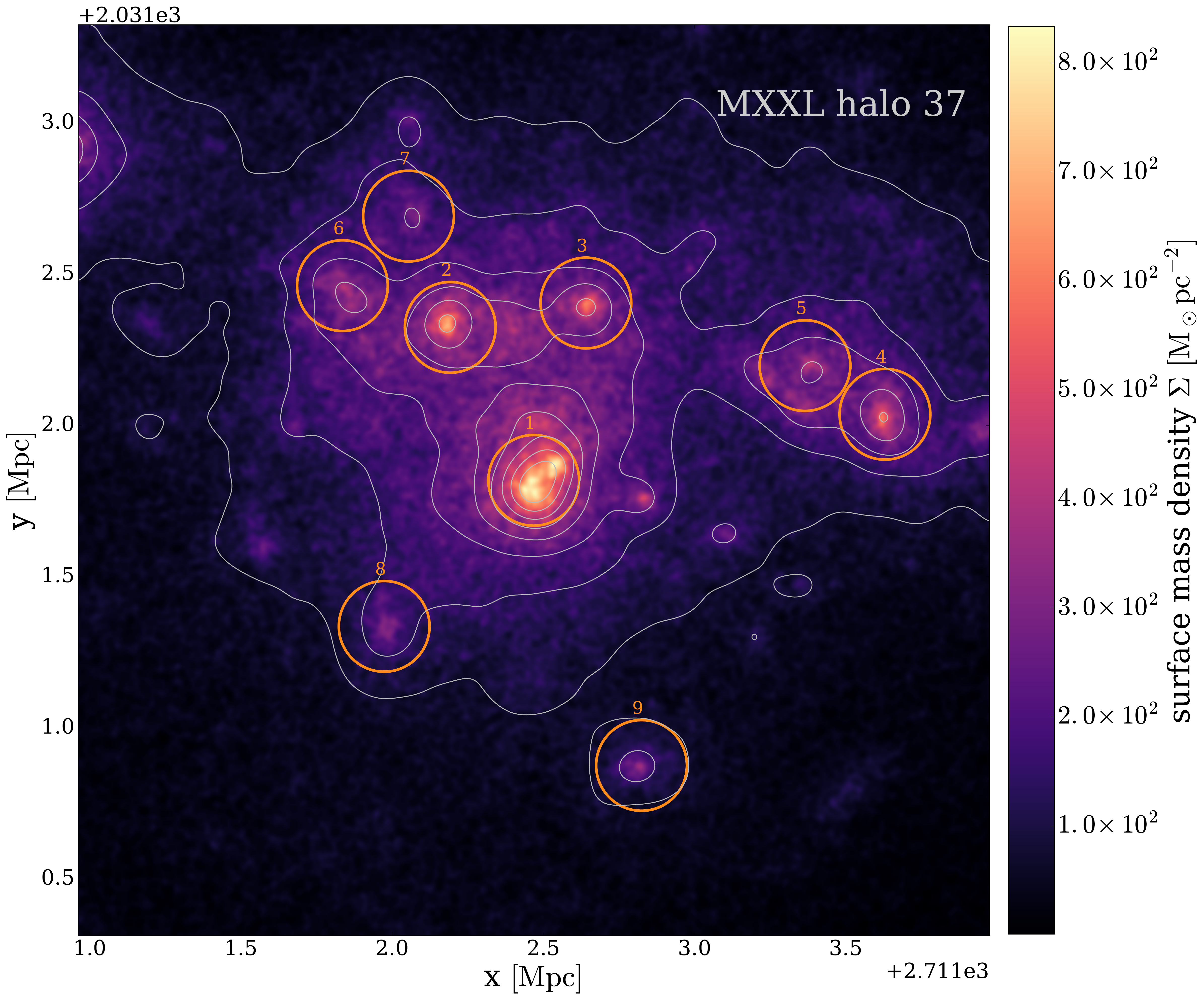}} &
\subfloat[]{\hspace{-0.9em}\includegraphics[width=0.45\linewidth]{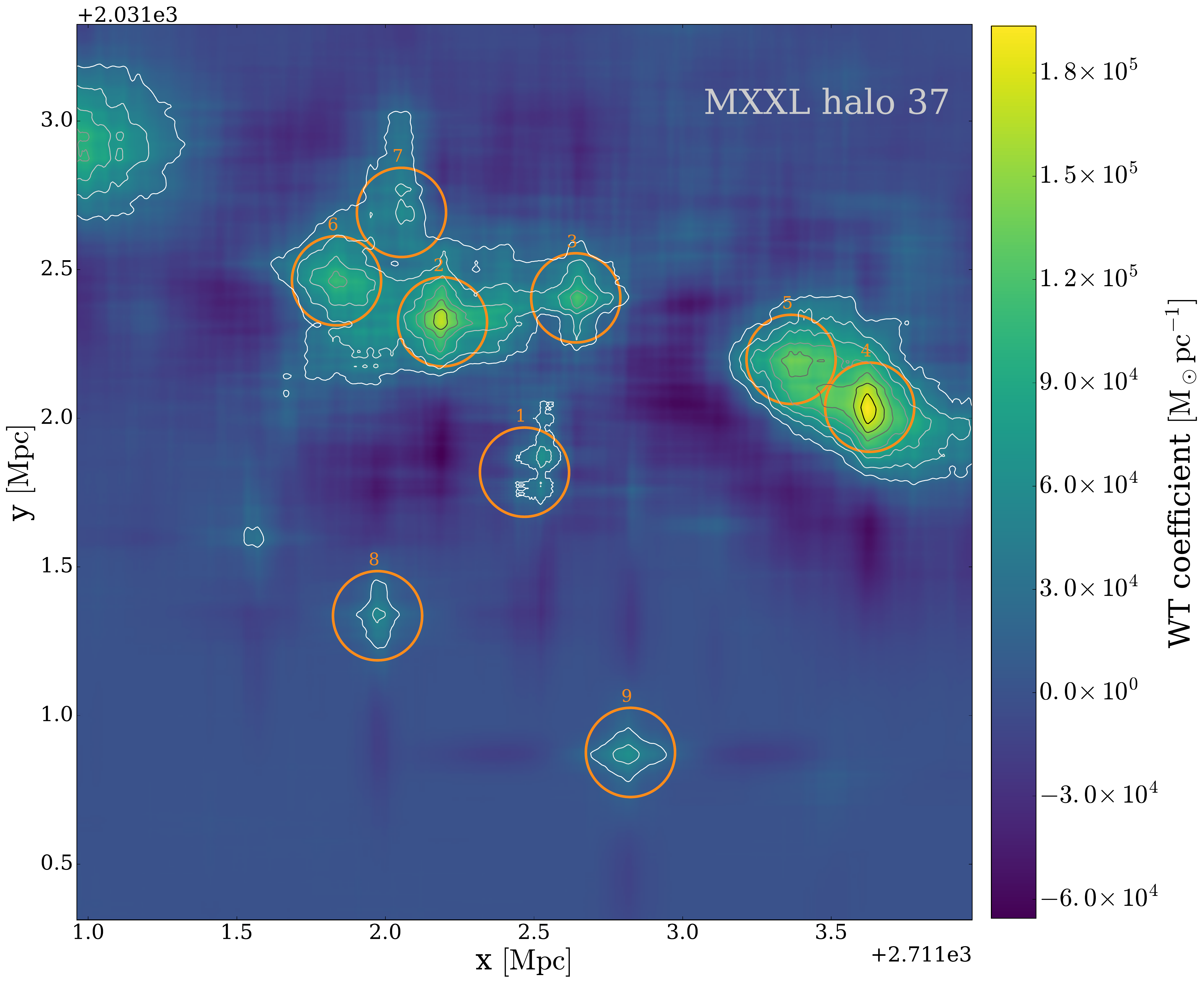}}\\
\subfloat[]{\hspace{-0.9em}\includegraphics[width=0.45\linewidth]{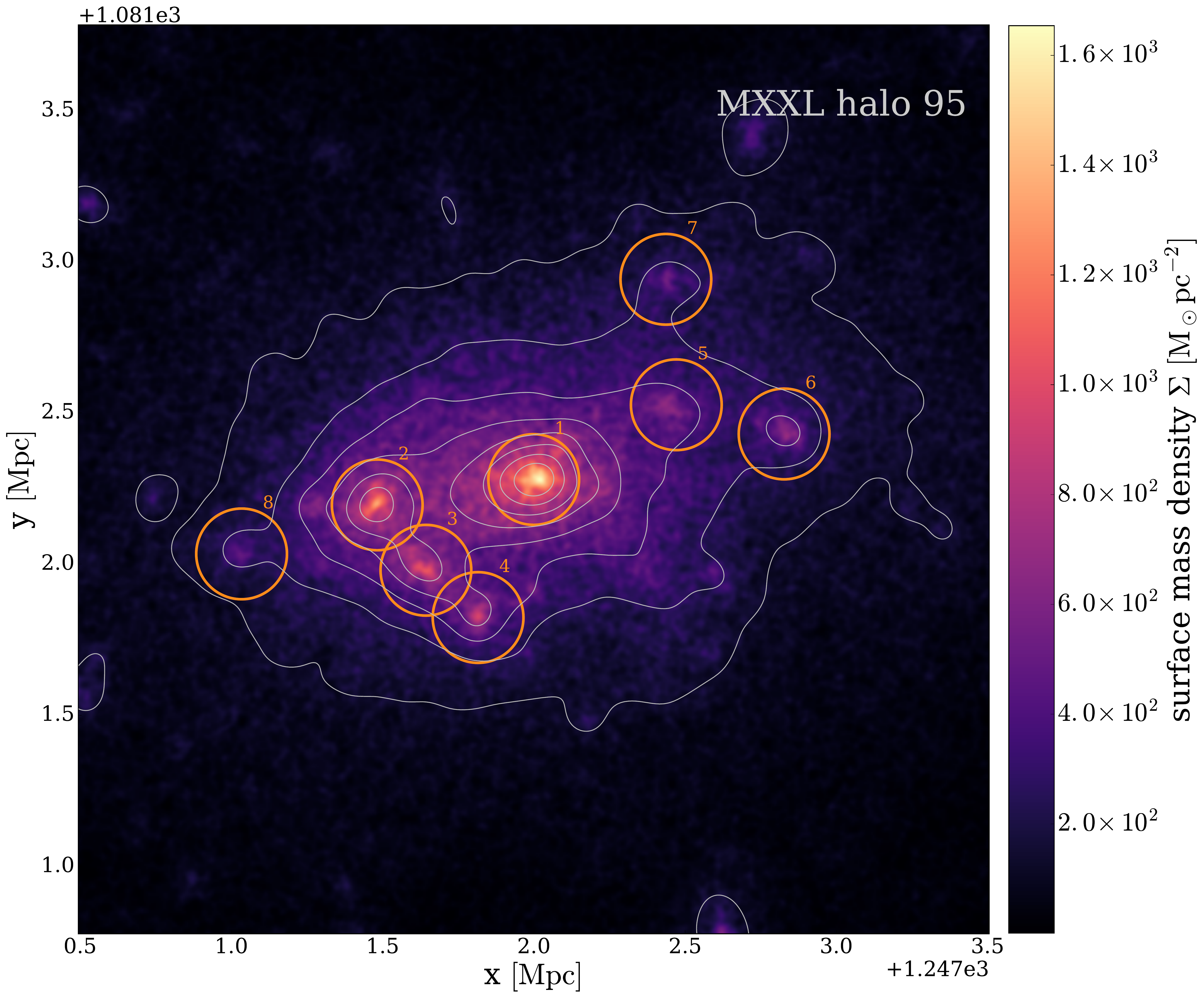}} &
\subfloat[]{\hspace{-0.9em}\includegraphics[width=0.45\linewidth]{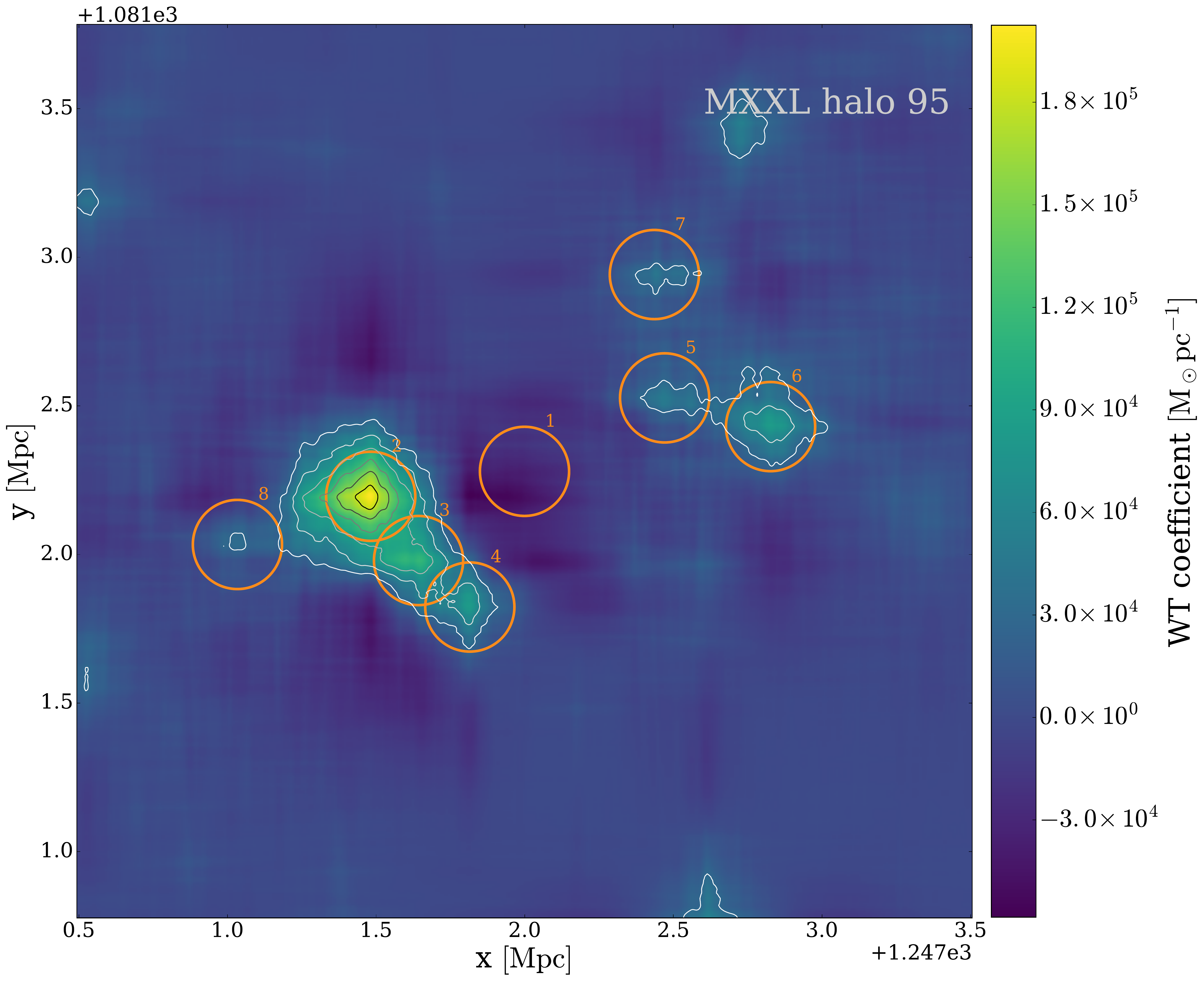}}\\
\subfloat[]{\hspace{-0.9em}\includegraphics[width=0.45\linewidth]{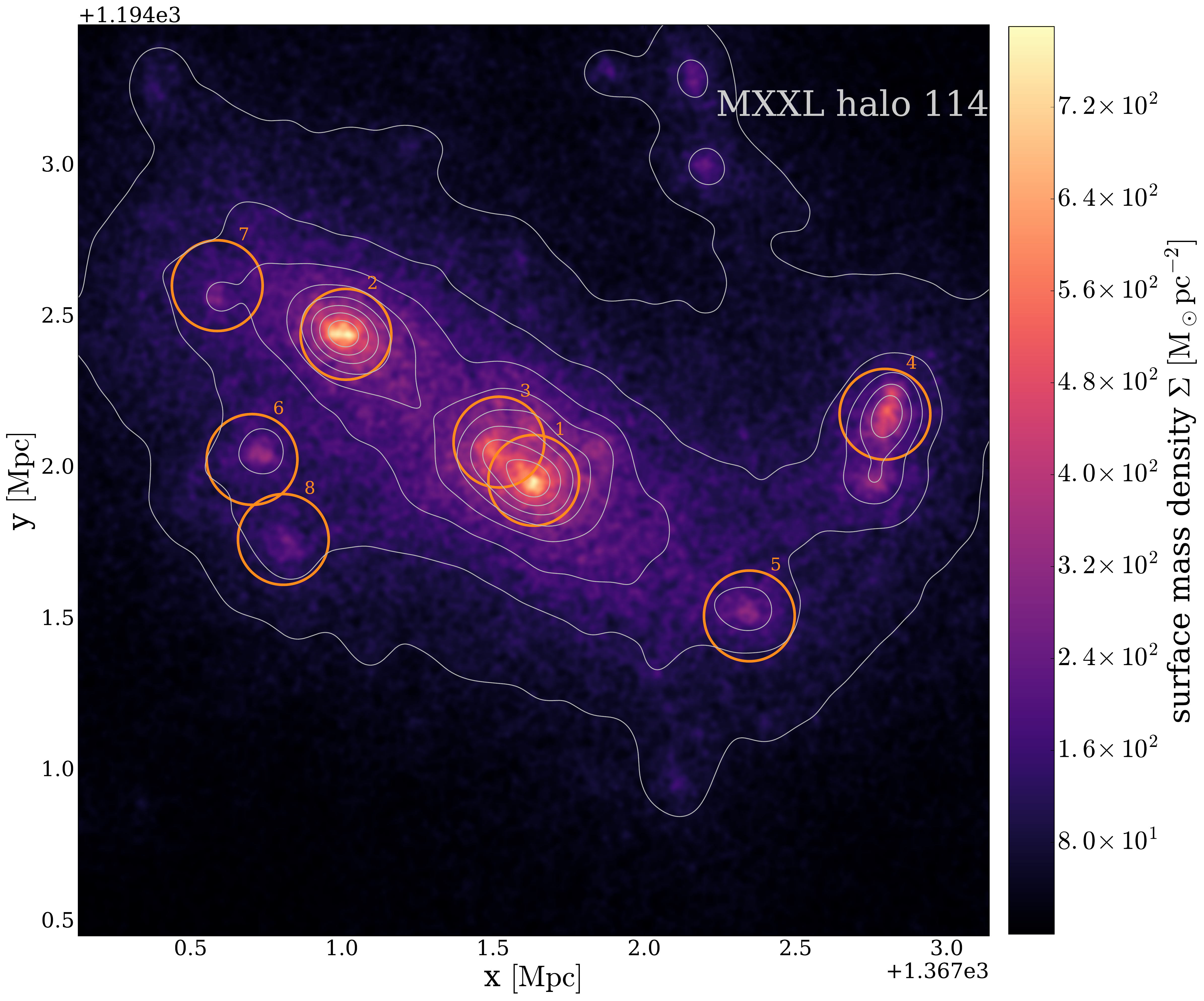}} &
\subfloat[]{\hspace{-0.9em}\includegraphics[width=0.45\linewidth]{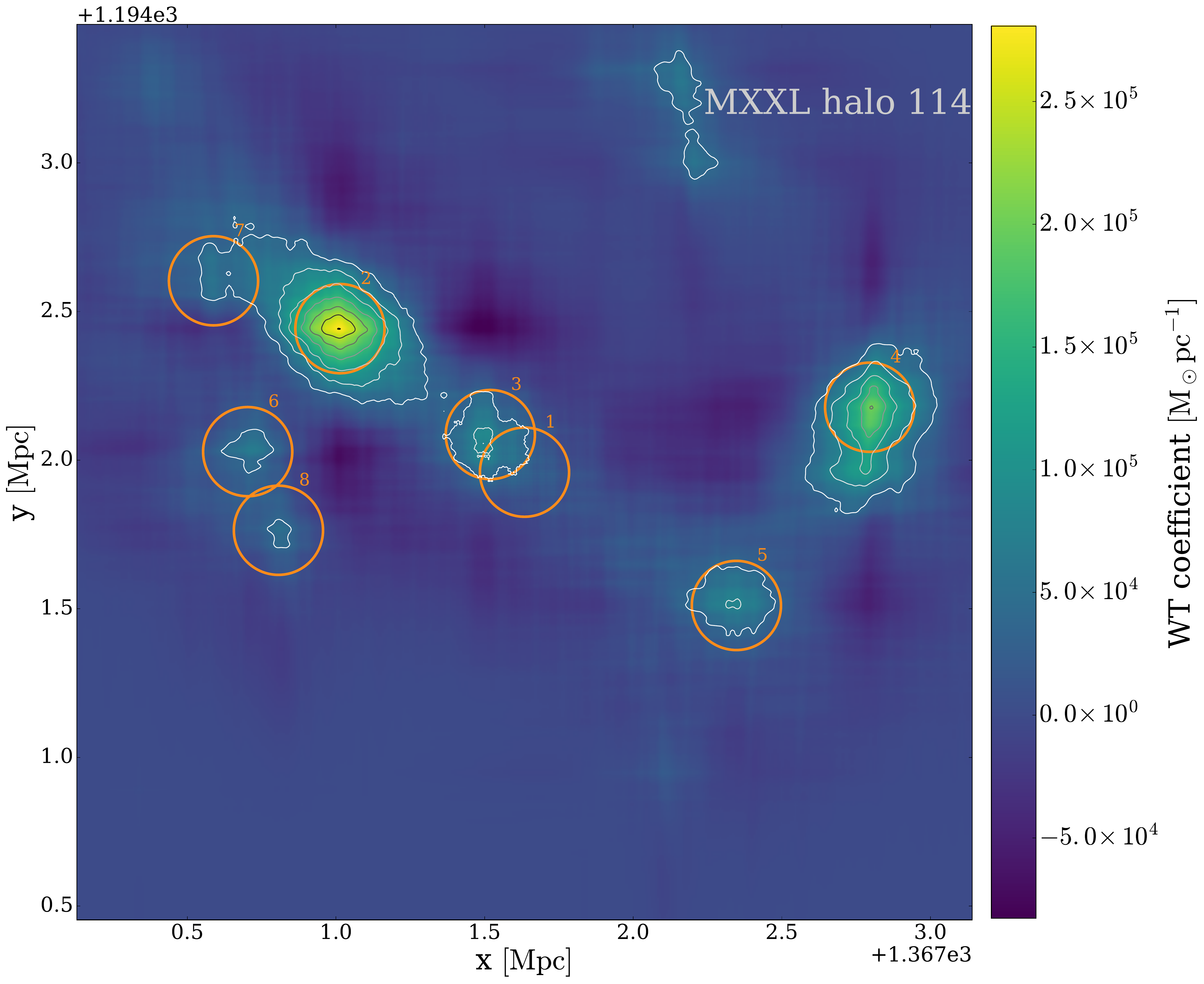}}\\\end{tabular}
    \caption{MXXL haloes 37, 95 and 114 which show a substructure distribution similar to that of Abell 2744. The panels on the \emph{left} side show the mass maps of all haloes where the colour map and contours show the surface density. The contours show the map after being smoothed by a Gaussian with standard deviation of 8 pixels ($\sim 36.4$~kpc). The \emph{right} panels show the wavelet transform coefficients used to identify the substructures. Substructures fulfilling the threshold criteria are marked as orange circles with radius corresponding to $R = 150$~kpc.}
    \label{fig:resultsSim1}
\end{figure*}%

\subsection{MXXL} 
\label{subsec:ResultsMXXL}
Using exactly the same method we search for Abell 2744-like haloes in the MXXL simulation. For this purpose we apply the wavelet transform algorithm to all mass maps obtained for the 209 MXXL haloes with a similar mass to Abell 2744 (as described in Section \ref{subsec:massMapsMXXL}). We find three MXXL haloes fulfilling all criteria with respect to the substructure distribution. The mass maps with highlighted substructures as well as the maps of the corresponding WT coefficients are shown in Fig.\,\ref{fig:resultsSim1}. The properties (i.e.\ aperture mass, distance from the centre and WT coefficient) of all substructures identified are listed in Table \ref{tab:substrComparison}.

The first halo (\emph{halo 37}) resembles the properties of Abell 2744 remarkably accurately. The cluster has a mass of $M_{37}(R<1.3\mathrm{Mpc}) = 2.61 \times 10^{15} \Msun$, similar to that of Abell 2744. Our wavelet transform algorithm identifies nine substructures within a projected distance of 1.2~Mpc. Sorting the subhaloes descending in their projected mass and comparing the mass of each rank to its equivalent in Abell 2744 shows a maximal discrepancy of 23 per cent between their aperture masses. Furthermore, the central substructure seems to consist of two separate density peaks, very similar to the bimodal mass distribution of the core of Abell 2744 \citep{Jauzac2015}. 

The second halo, \emph{halo 95}, has a mass of $M_{95}(R<1.3\mathrm{Mpc}) = 2.00 \times 10^{15} \Msun$ the least massive of the three MXXL haloes. We find eight substructures with an aperture mass higher than $3 \times 10^{13}\Msun$ within a distance of 1.0~Mpc from the centre. Comparing the substructures' aperture masses to those of Abell 2744 by sorting them in descending order of projected mass shows a discrepancy of at most 40 per cent. The higher discrepancy in comparison to that of \emph{halo 37} is due to the low masses of subhaloes \emph{6} to \emph{8}. These are slightly less massive than the least massive substructures of Abell 2744. The masses of the six most massive substructures differ by less than 16 per cent from those of Abell~2744. 

Finally, \emph{halo 114} has a mass of $M_{114}(R<1.3\mathrm{Mpc}) = 2.10 \times 10^{15} \Msun$, which is slightly lower than the mass of Abell~2744. Also this cluster consists of eight massive substructures within a radius of 1.25~Mpc from the centre. These lie in a mass range very similar to the substructures of Abell 2744 and differ by no more than 45 per cent. However, all substructures apart from the three most massive ones contain less mass than the substructures of Abell 2744.

These results depend only weakly on the choice of removing the main halo's mass profile approximated by an NFW-profile. In case of halo 37, subtracting the main halo's density profile has no effect at all. In case of halo 95 and 114, we find in each case one additional substructure if the NFW-profile is not subtracted. However, these additional substructures can hardly be detected by eye and it seems more likely that the background contribution of the main halo boosted the WT coefficients. In case of halo 114, substructure \emph{3} located in the cluster core can only be detected when the main halo is subtracted. However, the choice to subtract the main halo's mass or not has no influence on the detection of all clear substructures. The fact that removing the mass of the main halo has such a weak impact, shows clearly the desired behaviour to filter out only structures on subhalo scale while neglecting contributions by larger or smaller structures.

It should be noted that the WT coefficients of almost all substructures in the MXXL simulation are considerably higher than those of Abell 2744. These values could be slightly overestimated due to the finite mass resolution of the MXXL simulation. Since the dark matter distribution of the MXXL simulation is traced by particles of mass $m_{p} = 8.80 \times 10^{9}\, {\rm M}_{\odot}$, the cluster mass maps are not as smooth as in the observational case. As described in Section~\ref{subsec:massMapsMXXL}, we correct for this effect gently without removing substructures through smoothing by applying a Gaussian filter with a standard deviation of 1.5 pixels to each mass map from the MXXL simulation.

We furthermore use the mass maps to draw conclusions about Abell 2744's virial mass. In \cite{Schwinn2017}, the virial mass was predicted to be $M_{200} = 3.3 \pm 0.2 \times 10^{15} \Msun$ by using the projection of a corresponding NFW-profile. We compare this prediction with the $M_{200}$ masses of the three MXXL haloes investigated above. The first halo (\emph{halo 37}) has a mass of $M_{200,37} = 3.67 \times 10^{15} \Msun$, which is 40 per cent higher than its aperture mass within 1.3~Mpc. This agrees well with the prediction of \cite{Schwinn2017}. However, the masses of both of the other clusters, $M_{200,95} = 2.55 \times 10^{15} \Msun$ and $M_{200,114} = 2.41 \times 10^{15} \Msun$, are 11 per cent and 12 per cent, respectively, lower than expected from an extrapolation using an NFW profile. Since all of these clusters are undergoing a merger, the cluster is far from being relaxed, which explains the deviation from the extrapolation using an NFW-profile.

\subsection{Time evolution and projection effects} 
An inevitable shortcoming of our work is the analysis of the simulation data at redshift $z = 0.24$, while Abell 2744 is located at $z = 0.306$. This gap in the cluster evolution cannot be avoided, since the particle data of the MXXL simulation is not available for the redshift of Abell 2744. The only possibility to investigate the behaviour of the substructures in between these redshifts, is to trace the substructures back in time using the merger trees available for the {\tt SUBFIND} haloes. For each substructure found by the wavelet transform algorithm, we identify the closest {\tt SUBFIND} halo. In cases with more than one possible candidate, we select the most massive subhalo. This allows us to predict the positions and masses of the substructures at $z = 0.3$. We find that all substructures identified by the WT method in haloes \emph{37}, \emph{95} and \emph{halo 114} have corresponding {\tt SUBFIND} haloes.

\subsubsection{Change of distance during the infall} 
\begin{figure}
\includegraphics[width=\linewidth]{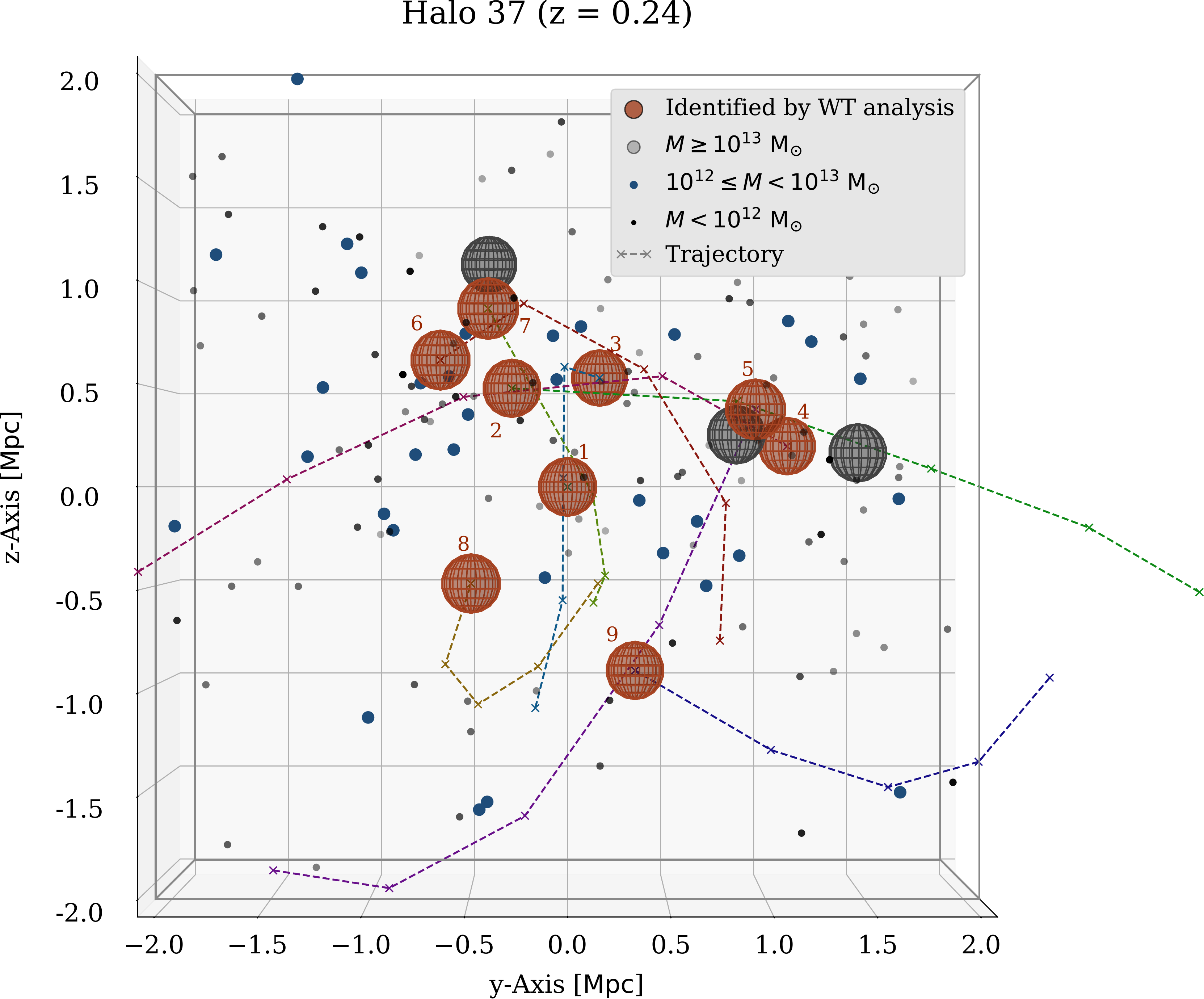}
\caption{Time evolution of the substructures in MXXL \emph{halo 37}. The {\tt SUBFIND}-haloes corresponding to the nine substructures found by the WT algorithm are shown as red spheres with $R  = 150$\,kpc. Their trajectories between snapshot 50 ($z = 0.41$) and 54 ($z = 0.24$) are shown as dashed lines. All other haloes are shown depending on their mass either as grey spheres, blue or black dots.}
\label{fig:timeEvolution}
\end{figure}

The time evolution of the {\tt SUBFIND} haloes corresponding to the substructures found by the WT analysis is shown in Fig.\,\ref{fig:timeEvolution} for MXXL \emph{halo 37}. The trajectories of the infalling substructures are heavily affected by the ongoing merger and move up to 1.9~Mpc between two snapshots. However, interpolating the trajectories shows that at least seven of the identified substructures are within a radius of 1.2~Mpc from the main halo at $z = 0.3$. In case of \emph{halo 95}, the trajectories of the eight identified subhaloes describe a merger similar to that of \emph{halo 37}. At $z = 0.3$, 6 of the 8 {\tt SUBFIND} haloes are already within a radius of 1.2~Mpc. In contrast to the other haloes, the subhaloes of \emph{halo 114} do not have as perturbed trajectories, but they simply fall into the cluster. However, at $z = 0.3$ only 4 of the 8 substructures are within a radius of 1.2~Mpc from the cluster centre.

\begin{figure*}
\begin{tabular}{cc}
\subfloat[]{\hspace{-1em}\includegraphics[width=0.45\linewidth]{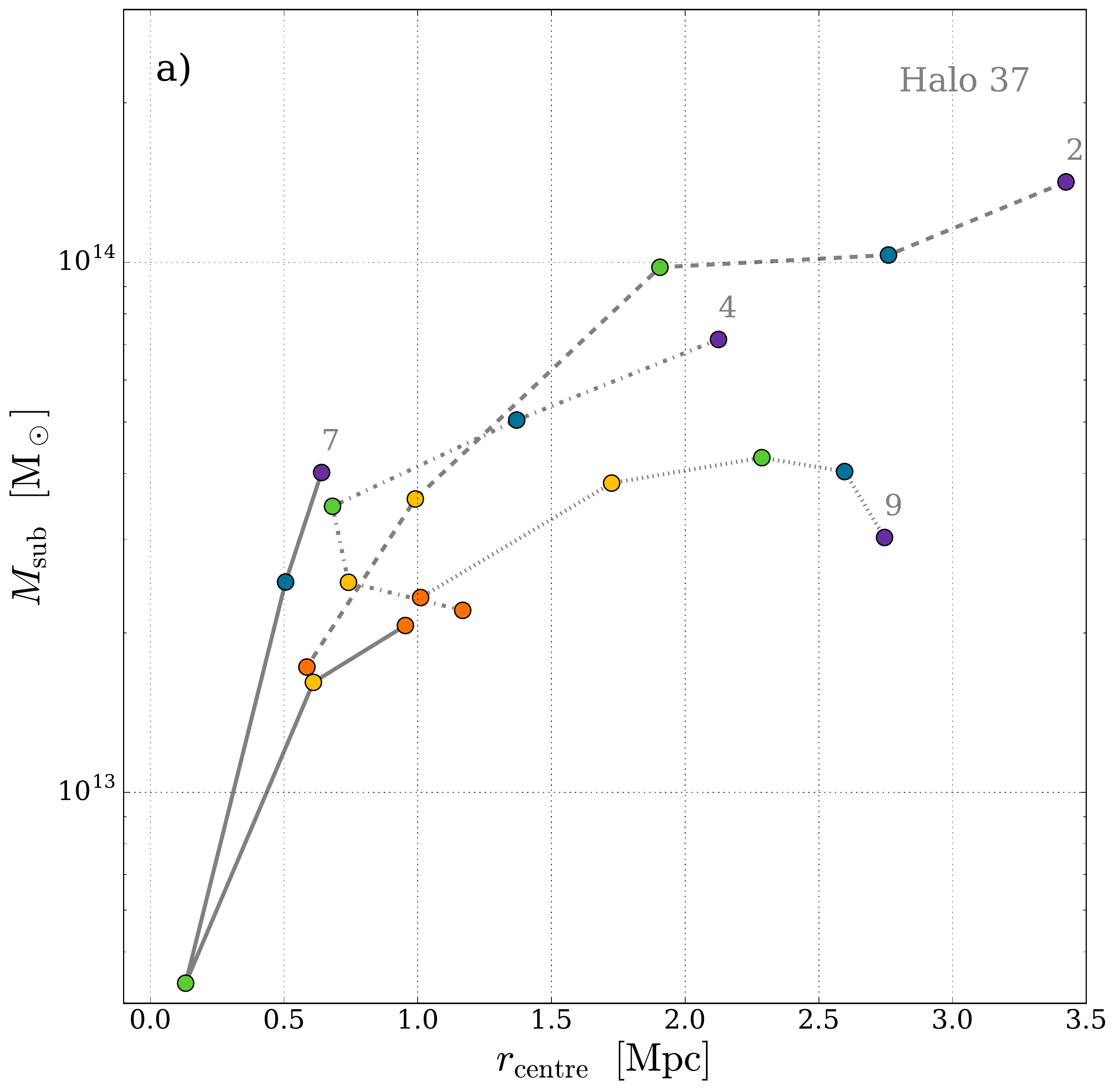}} &
\subfloat[]{\hspace{3.5em}\includegraphics[width=0.45\linewidth]{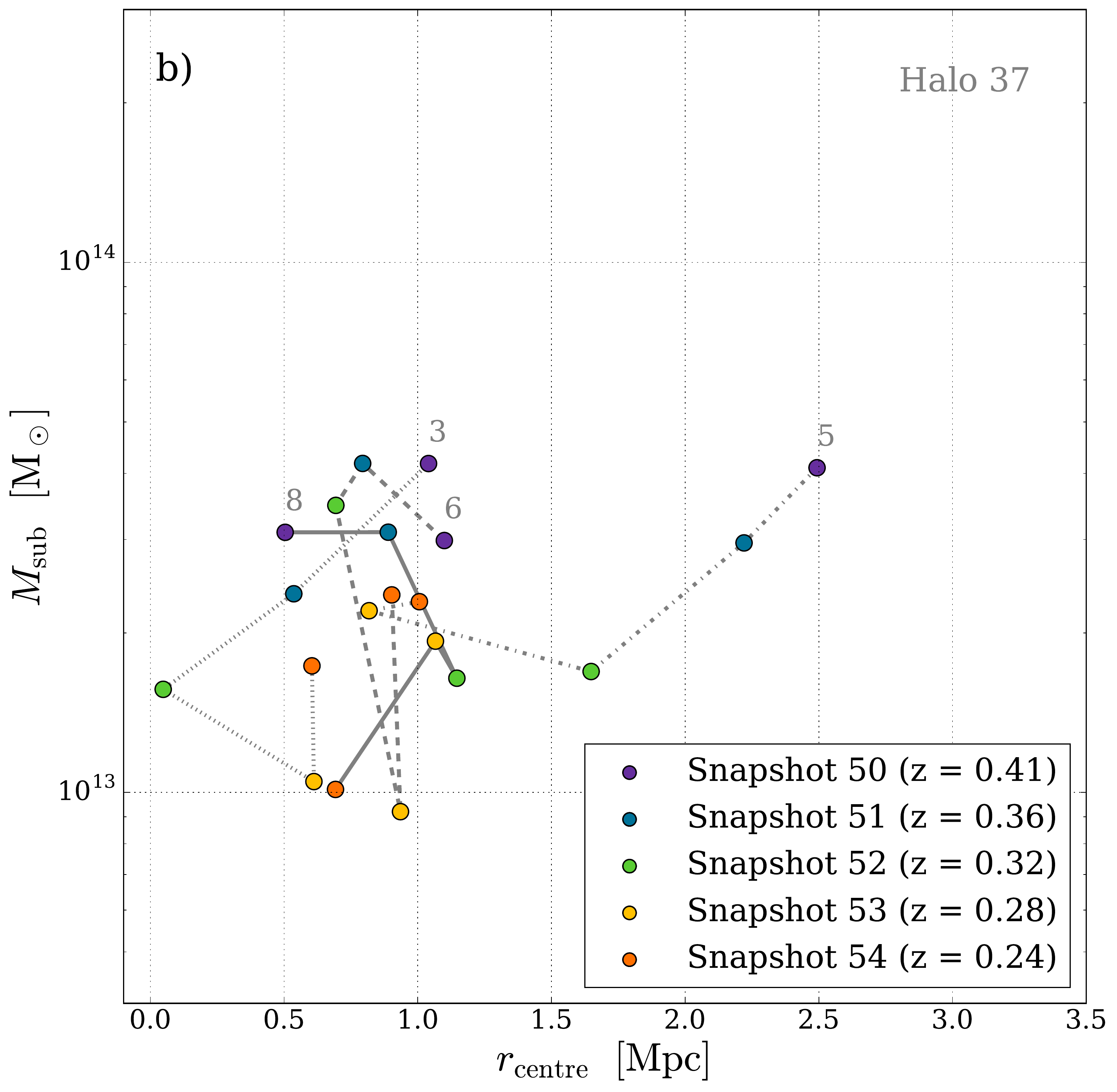}}
\end{tabular}
    \caption{Time evolution of the mass and radial distance of the subhaloes of \emph{halo 37}. The \emph{left} panel shows subhaloes \emph{2}, \emph{4}, \emph{7} and \emph{9}. The \emph{right} panel shows subhaloes \emph{3}, \emph{5}, \emph{6} and \emph{8}. The five different snapshots (50-54) are colour coded as given in the legend. The radius is given as the distance from the position of the central halo.}
    \label{fig:substructurePaths}
\end{figure*}%
\subsubsection{Change of mass during the infall}
We also investigated the mass evolution of \emph{halo 37}'s subhaloes over the five preceding snapshots (i.e.\ between $z = 0.41$ and 0.24), shown in Fig.\,\ref{fig:substructurePaths}. For clarity, we plot two panels each showing the evolution of four subhaloes. The subhaloes show a mixture of mass growth and mass being stripped away due to the infall. While some substructures (subhaloes 2, 4 and 9) are purely affected by tidal stripping during their infall, others (i.e.\ subhaloes 3, 5, 6, 7 and 8) also gain mass between one or two snapshots. The mass evolution of subhalo 7 reveals clearly the difficulties faced by {\tt SUBFIND} to identify subhaloes that are very close to the cluster centre. It loses 90 per cent of its mass between snapshot 50 and 52 when it is very close to the centre. Being more distant again in snapshot 53, its mass is restored from 10 per cent back to 40 per cent of its mass at snapshot 50. This emphasises the problems of analysing masses of substructures close to the centre based on the {\tt SUBFIND} data. The high fluctuation in mass of the substructures makes it difficult to predict the mass change of the substructures between $z = 0.3$ and $z = 0.24$. However, the mass of 4 subhaloes decreased from $z = 0.3$ to 0.24, the masses of the central halo and subhalo 6 increased only by 10 percent and 7 percent, respectively. Furthermore, the projected mass of subhalo 7 is not very likely to be affected as dramatically as the change in {\tt SUBFIND}-mass suggests. Since it is much closer to the centre at $z=0.3$, it is more likely that the projected aperture mass is even higher due to the additional contribution of the host halo. We therefore argue that it is likely to find at least seven substructures with similar aperture masses in \emph{halo 37} at redshift $z = 0.3$.

\subsubsection{Projection effects}
The identification of the corresponding {\tt SUBFIND} haloes allows us to investigate the distribution of the subhaloes along the line-of-sight. Taking the full 3D information into account shows that the identified substructures are distributed over a distance of almost 4~Mpc centred on the main halo. As expected, subhaloes appearing close to each other on the map can be quite distant. While subhaloes {\it 2}, {\it 3}, {\it 6} and {\it 7} form a group on the 2D map, they are distributed over a line-of-sight distance of 3.4~Mpc. However, another projection effect -- the addition of multiple subhaloes within the aperture -- has only a minor effect. As Fig.\,\ref{fig:timeEvolution} shows, only subhalo {\it 5} is affected by another close massive subhalo which appears as separate peak in the mass map in Fig.\,\ref{fig:resultsSim1}. 

\section{Masses of subtructures}
\label{sec:masses}
\subsection{Comparison of aperture masses with SUBFIND masses}
The first analysis based on {\tt SUBFIND} haloes in \cite{Schwinn2017} led to finding considerably lower substructure masses than observed in Abell 2744. As \cite{Mao2017} point out, this is partly due to additional mass in the body of the halo being projected onto the aperture. Using the {\tt SUBFIND} haloes found for \emph{halo 37}, we compare the {\tt SUBFIND} masses and the corresponding aperture masses for all substructures apart from the central halo. The results are listed in Table~\ref{tab:masses}. Similar to the analysis in \cite{Mao2017}, we find that the aperture mass of the substructures can reach values up to five times their corresponding {\tt SUBFIND} mass due to line-of-sight projection of additional mass in the main halo. 

We try to estimate if this effect can be included in an analysis based on {\tt SUBFIND} data only as in \cite{Schwinn2017}, which would help to analyse simulations where the full particle data are not available for all snapshots. We therefore integrate the mass within a cylinder of length 30~Mpc and radius 150~kpc provided by two NFW profiles -- the first modelling the main halo with $M_{200}  = 3.67 \times 10^{15}\Msun$ and the second modelling the subhalo while estimating the $M_{200}$-mass by $M_\mathrm{sub}$. In this integration, we adopt the $c$-$M_{200}$ relation presented in \cite{Neto2007}. We show in Table~\ref{tab:masses}, column 5 the expected contribution of the host halo. This estimate shows that the host halo can contribute between 23 and 71 per cent of the aperture masses measured from the mass maps directly (column 3). The expected total masses (i.e.\ the contributions of host halo and subhalo combined) are shown in column 6. The extrapolated masses of all subhaloes apart from subhaloes 8 and 9 are considerably lower than the actual masses measured from the mass map directly. The extrapolation underestimates the projected mass by up to 60 per cent. This shows that the translation from subhalo masses measured with {\tt SUBFIND} to projected masses within a 2D mass map is not possible. This is most likely due to the fact that the cluster is far from being relaxed and thus assuming spherical symmetry and an NFW-profile leads to incorrect results. Additionally an overprediction of tidal stripping by {\tt SUBFIND} as reported in \cite{Muldrew2011} and \cite{Han2017} would as well lead to lower expected masses. It is therefore best to use masses obtained from the particle data directly.

We furthermore compare the subhalo masses obtained by {\tt SUBFIND} (column 4) to those obtained in the mass map. This comparison shows that the {\tt SUBFIND} mass can be up to 80 per cent lower than the mass measured within an 150~kpc-aperture.

\begin{table}
  \caption{Comparison of different mass estimates for all substructures of \emph{halo 37} apart from the central halo. The table lists subhalo ID (column 1), distance from the central halo (column 2), the mass measured within a 150~kpc aperture from the projected mass map (column 3), the mass provided by {\tt SUBFIND} for the closest {\tt SUBFIND}-halo (column 4), the mass of the host halo at the position of the substructure estimated assuming an NFW-profile (column 5) and the expected aperture mass assuming two NFW-profiles for the main halo and the {\tt SUBFIND} subhalo (column~6).}
  \vspace{0.5em}
  \centering

  \resizebox{\linewidth}{!}{%
  \begin{tabular}{ccccccc}
    \hline\hline
    \noalign{\smallskip}

    {\it ID} & $D_{C-S}$ & $M ( r < 150 \, {\rm kpc} ) $ & $M_\mathrm{sub}$ & $M_\mathrm{host}$ & $M_\mathrm{extr}$ \\\noalign{\smallskip}  
             & (kpc)  & ($10^{13} {\rm M}_\odot$) & ($10^{13} {\rm M}_\odot$) & ($10^{13} {\rm M}_\odot$) & ($10^{13} {\rm M}_\odot$) \\\noalign{\smallskip} \hline 
    \noalign{\smallskip} 
2 & 576 & 8.85 & 1.72 & 3.80 & 4.79  \\[0.5em]
3 & 615 & 7.28 & 1.73 & 3.35 & 4.35 \\[0.5em]
4 & 1182 & 6.97 & 2.21 & 1.61 & 2.80 \\[0.5em]
5 & 974 & 6.73 & 2.29 & 2.13 & 3.36 \\[0.5em]
6 & 902 & 6.37 & 2.36 & 2.35 & 3.60 \\[0.5em]
7 & 967 & 5.00 & 2.06 & 2.19 & 3.33 \\[0.5em]
8 & 683 & 4.08 & 1.01 & 2.88 & 3.54 \\[0.5em]
9 & 1008 & 3.08 & 2.33 & 2.03 & 3.27 \\[0.5em] 
  \hline\hline
  \end{tabular}
  }
  \label{tab:masses}
\end{table}%
\subsection{Substructure finding based on SUBFIND}
In a recent study by \cite{Mao2017}, the particle data of the Phoenix set of very high resolution simulations \citep{Gao2012} was investigated at $z=0.32$ to search for haloes similar to Abell 2744. To do so they analysed the substructures of the most massive halo in the simulation, which is the only halo as massive as Abell 2744. They computed the aperture mass for all subhaloes with $M_\mathrm{sub} \geq 2.3 \times 10^{11}\Msun$ for 24 different projections of the halo. By doing so they found at least three projections with eight and another one with nine substructures.

Our first analysis presented in \cite{Schwinn2017} led to the conclusion that there is no halo in the MXXL simulation with eight or more subhaloes as massive and as close as in Abell 2744 when only the {\tt SUBFIND} results are used. Our method did not include the contribution of the matter distribution of the main halo, since there was no particle data available. 
In contrast to that, the method used by \cite{Mao2017} could exaggerate the influence of the host halo on the subhalo apertures due to their method based on aperture masses alone. As they state correctly, their method does not ensure that a halo found by {\tt SUBFIND} is actually significant enough to be detected as a substructure in a weak lensing mass map. Since the host halo contributes such a large fraction to the total mass, \cite{Mao2017} are prone to picking up light subhaloes while the necessary aperture mass is provided mainly by the diffuse host halo mass. However, such a subhalo would not correspond to the substructures found in the observation of Abell 2744. Although, our wavelet based method detects a few light subhaloes as well (e.g.\ subhaloes \emph{8} in \emph{halo 95} and \emph{halo 114}) it is guaranteed that these substructures are detected in the observed and simulated data sets in the same way. Furthermore, using only {\tt SUBFIND} haloes could potentially lead to missing significant substructures in the mass map. This would be the case if apparent substructures are the result of line-of-sight projection. In this case the substructures would not have a {\tt SUBFIND} counterpart. However, this was not the case for the MXXL haloes investigated in this study. It seems thus advisable to use -- for this kind of comparison of observations with simulations -- a method that treats both data sets equally.

\subsection{Artificial disruption of substructure}
When investigating subhaloes in N-body simulations, it is important to keep in mind that not only their identification but also their evolution in the simulation itself can be affected by numerical processes. A very detailed investigation of the influence of parameter choices in the N-body simulation on the tidal disruption of subhaloes has been performed recently by \cite{vandenBosch2018}. Using the simplified setting of a subhalo on a circular orbit in a static, analytic host halo, they addressed the question of whether the tidal disruption of subhaloes has a physical or numerical origin. They find that mainly two effects have an important influence and cause a spurious disruption of subhaloes.
  
The first is due to the force softening parameter which is commonly introduced by hand in N-body simulations. It is set to prevent the gravitational potential between two particles from diverging when two particles approach each other. For this reason a minimal distance $\epsilon$ is added quadratically to the separation of the two particles. Several studies exist on the optimal choice of this parameter \citep{vanKampen2000,Dehnen2001,Power2003}. However, \cite{vandenBosch2018} find that the commonly chosen values lead to a spurious disruption of subhaloes on orbits close to the centre. 

The second effect is the amplification of discreteness noise by a runaway instability.  Since the subhalo is represented by discrete particles there exist different equivalent realisations of the same subhalo. If one the realisations loses more mass through tidal stripping than the average, it expands more than average due to revirialisation. Thus, again more particles than average are beyond the subhalo's virial radius and get stripped away. This runaway instability leads to a large variance of the time it takes to disrupt the subhalo.

These effects also have the potential to influence the findings of our work. \cite{vandenBosch2018} show drastic numerical effects for orbits close to the host halo centre (i.e.\ $R_\mathrm{orb} = 0.1 R_\mathrm{vir}$) for the later stages of the infall where the subhalo has already lost more than 90 per cent of its original mass. The majority of subhaloes considered in our work are on orbits with $R_\mathrm{orb}  > 400$~kpc. Since Abell 2744 has a virial radius of 2.8~Mpc, this corresponds to $R_\mathrm{orb}  > 0.2 R_{200}$. For these larger orbits, \cite{vandenBosch2018} show that the numerical bias is still present, but less drastic. This is true especially for the earlier phases of the infall where 10 per cent of the original subhalo mass is still gravitationally bound to the subhalo. In order to quantify their findings, \cite{vandenBosch2018} give two equations to evaluate up to which bound mass fractions stripped subhaloes can be deemed trustworthy (their eq. 20 and 21). We evaluate these equations with the parameters of the MXXL simulation, i.e.\ particle mass $m_\mathrm{p} = 8.80 \times 10^9 \mathrm{M}_\odot$ and softening length $\epsilon = 13.7$~kpc, and estimate the concentration of the subhaloes with the $c$-$M$ relation of \cite{Neto2007}. This allows us to assess if the substructures identified in the MXXL haloes are significantly influenced by numerical effects.  

We find that if infalling subhaloes had an original mass of $M_\mathrm{orig} = 10^{14} \mathrm{M}_\odot$, they are not significantly affected by numerical processes until they are stripped down to a mass of $\sim 5 \times 10^{12} \mathrm{M}_\odot$. Even if the subhalo's original mass is much higher ($M_\mathrm{orig} = 10^{15} \mathrm{M}_\odot$), numerical effects begin to influence its disruption significantly when the remaining mass is below $\sim 7.5 \times 10^{12} \mathrm{M}_\odot$. Since the substructures we are analysing have not been stripped to this extend, i.e.\ their masses are considerably higher than these lower thresholds, we may assume that they are not significantly affected by numerical processes. However, it is worth having in mind on the basis of this discussion that substructures in the close centre of clusters are by no means exact representations of the true mass distribution. It leaves the possibility that the subhalo masses in our analysis may be slightly underpredicted even if they are determined directly from the mass maps. It is therefore possible that in a simulation with smaller numerical effect, we would find even more substructures fulfilling our criteria.

\section{Summary}
\label{sec:summary}
We have searched the MXXL simulation for haloes with properties similar to the galaxy cluster Abell 2744. Our analysis is built upon our findings in \cite{Schwinn2017}, where we used the FoF- and {\tt SUBFIND} data only. These findings suggested a potential tension between the observations of Abell 2744 and the predictions of $\Lambda$CDM. Here we have used the particle data of the MXXL stored at $z = 0.24$ in order to treat observational and simulated data as similar as possible during the comparison.
We selected haloes with a total mass similar to Abell 2744 and produced projected mass maps comparable to that obtained for Abell 2744. We then investigated the substructures within these haloes according to three criteria: $i)$ their distance from the centre, $ii)$ their projected mass within an aperture of 150~kpc radius and $iii)$ their significance in the mass map. In order to detect the substructures and to quantify their significance, we used a wavelet transform (WT) algorithm. This algorithm allowed us to filter the mass map at a scale of 182~kpc and to define a significance threshold. With this to hand, we searched for a cluster with seven substructures fulfilling the following criteria: 

\begin{itemize}[leftmargin=1em]
\setlength{\itemindent}{-0.5em}
\item a distance less than 1.25~Mpc from the cluster centre,\\[-0.8em]
\item an aperture mass of $M(R<150 \mathrm{kpc}) \geq 3 \times 10^{13}\,\Msun$ and\\[-0.8em]
\item a WT coefficient of $W \geq 2.6 \times 10^{10}\,\Msun \mathrm{pc}^{-1}$ at a scale of 182~kpc\footnote{This value restores as many of the substructures found by \cite{Jauzac2016} as possible.}.
\end{itemize}

We found three haloes within the MXXL simulation with a substructure distribution similar to Abell 2744. The probability of a cluster like Abell 2744 to be observed in a $\Lambda$CDM universe can be estimated very roughly by comparing the simulation volume to that of the sphere up to Abell 2744's redshift (z=0.306). Since the simulation volume is ten times bigger than the volume out to z=0.306 and we find three similar clusters, the probability of finding Abell 2744 can be estimated to be approximately 30 per cent. This, however, does not take into account that we analyse the particle data of the simulation at $z=0.24$ and thus it can only serve as a rough estimate. It shows qualitatively that, albeit being a rare object, Abell 2744 is not in tension with a $\Lambda$CDM universe. This result resolves the discrepancy reported in \cite{Schwinn2017} after analysing the FoF and {\tt SUBFIND} data only. While investigating the reason for this divergence between the different data sets, we find that the host halo contributes a high amount to the mass measured in the 150~kpc aperture. 
Due to the unrelaxed state of the clusters, this additional mass cannot simply be added to the {\tt SUBFIND} mass of the subhaloes, e.g. by assuming an NFW-profile for the main halo. Although {\tt SUBFIND} is still one of the most reliable subhalo finders available for simulations \citep{Behroozi2015}, it seems therefore more advisable to use a method that can be equally applied to mass maps from gravitational lensing in case of a comparison as has been presented here. For such a method, it is necessary to create projected mass maps from the particle data of the simulation which can then be analysed in the same fashion as the mass maps obtained for observed clusters.

However, the particle data are not available at the redshift of Abell 2744 in the MXXL simulation. To close the gap between Abell 2744's redshift $z=0.3$ and $z=0.24$ at which the particle data of the MXXL simulation are available, we used the {\tt SUBFIND} merger trees and trace the substructures back in time. For each substructure found with the WT algorithm we identified the closest {\tt SUBFIND} halo and investigated its mass and distance to the central subhalo at Abell 2744's redshift ($z=0.306$). Although some of the substructures experience drastic changes in mass and distance due to the dynamical state of the clusters, we find that it is reasonable to assume that at least seven substructures can be found close to the centre for at least one of the haloes. By applying the criteria proposed by \cite{vandenBosch2018}, we showed that numerical effects leading to the spurious disruption of subhaloes do not have a significant impact on this result.

We have introduced a robust approach to finding substructures in observed and simulated clusters equally and have demonstrated its usefulness by applying it to study Abell 2744 as a proof of concept. It will be instructive to apply our approach to other massive clusters with substantial substructure, e.g.\  the \emph{Hubble Frontier Field} clusters \citep{Lotz2016} and ``El Gordo'' \citep{Marriage2011,Menanteau2012}.

\section*{Acknowledgements}
The authors are grateful to Joop Schaye and the anonymous referee for helpful comments and suggestions. Furthermore, we would like to thank Volker Springel for providing the particle data of snapshot 54 of the MXXL simulation. 

JS thanks Daniele Sorini for helpful discussions and acknowledges financial support from the Heidelberg Graduate School of Fundamental
Physics (HGSFP). CMB acknowledges receipt of a Leverhulme Trust Research Fellowship. This work was supported by the Science and Technology Facilities Council [ST/L00075X/1]. We used the DiRAC Data Centric system at Durham University, operated by the Institute for Computational Cosmology on behalf of the STFC DiRAC HPC Facility (www.dirac.ac.uk). This equipment was funded by BIS National E-Infrastructure capital grant ST/K00042X/1, STFC capital grants ST/H008519/1 and ST/K00087X/1, STFC DiRAC Operations grant ST/K003267/1 and Durham University. DiRAC is part of the National E-Infrastructure.
\bibliographystyle{mnras}%
\bibliography{A2744_MassMaps}

\bsp	
\label{lastpage}
\end{document}